\documentclass[showpacs,preprintnumbers,amsmath,amssymb]{revtex4}

\begin{document}

\preprint{}

\begin{center}
{\bf \Large Random Energy Model with complex replica number,
complex temperatures and classification of the string's phases}\\
\vspace{4mm}
D.B. Saakian\\

Yerevan Physics Institute,
Alikhanian Brothers St. 2,\\ Yerevan 375036, Armenia
\\
\end{center}

\begin{abstract}
The results by E. Gardner and B.Derrida have been enlarged
for the complex temperatures and complex numbers of replicas.
The phase structure is found. There is a connection with string
models and their phase structure is analyzed from the REM's point  of view.
\end{abstract}
\maketitle

\section{Introduction}

Random Energy Model (REM) [1-5] is connected with the many problems of  
modern      
physics.
In [6-8] has been found, that correlators in the directed model are connected 
with the free energy in directed polymer. The  
last is equivalent to REM in thermodynamic limit. 

Liouville model is closely connected with the bosonic string in the d-dimensional
Euclidean space [9].
It is easy to check, that the connection of strings with REM is even stronger. If one 
considers the  integration of string's partition via area of closed surfaces,
 [10-13] then after integration via zero mode of a Laplacian 
 an expression is obtained for the partition like to REM with finite replica numbers 
(solved for real temperatures in [5]):
\begin{eqnarray}
\label{e1}
Z\sim
\int D_g\phi e^{\frac
{1}{8\pi}\int d^2w\sqrt{\hat g}{\phi \Delta \phi +QR\phi}}
(\int d^2w
\sqrt{\hat g}e^{\alpha\phi})^{-\frac{Q}{\alpha}}
\end{eqnarray}
Here  $\phi(w)$ is a field on closed 2-d surface,
$\alpha,Q$ are parameters real for  $d<1$, R is a curvature, $D_g\phi$ is a measure.
$Q,\alpha$ are defined by $d$ according to formulas of David-Distler-Kawai (see review [12]).
The analytical continuation of parameters $Q,\alpha$ at $ d>1$
gives complex value for parameters (see section 4). 

One can understand the last expression as an average of the 
$\mu=-\frac{Q}{\alpha}$-th degree 
(replicas number) of
the sum $\sum_ie^{\alpha\phi(w_i)}$ via normal distribution of variables
 $\phi_i\equiv \phi(w_i)$ with a quadratic form
\begin{eqnarray}
\label{e2}
\frac{1}{8\pi}\int d^2w\sqrt{\hat g}
{\phi \Delta \phi +QR\phi}
\end{eqnarray}
The main idea of this work (following to [6-8]) is that the phase structure of the (1)
 can be mapped to other models with the simpler choice of quadratic form in the exponent of normal
distribution that the one in Eq. (2). We are going to connect 
the system (1) with the chain of models (each one with the same phase structure as the previous 
one but simpler), where the last one in the chain is the Random Energy Model (REM).
That's why we decided to solve REM at complex temperatures [13-15] and complex replicas numbers.  \\
REM is a model with   
\begin{equation}
\label{e3}
P(E)=\frac{1}{\sqrt{\pi 2N}}\exp[-\frac{E^2}{2N}]
\end{equation}
The total distribution of energies is factorized: for 
$1\le \alpha\ne\beta\le M$
\begin{equation}
\label{e4}
P(E_{\alpha},E_{\beta})=P(E_{\alpha})P(E_{\beta})
\end{equation}
The main our interest is connected with partition
\begin{eqnarray}
\label{e5}
z=\sum_i\exp\{-\beta E_i\}\nonumber\\
Z=<z^{\mu}>,
\end{eqnarray}
for a general value of $\mu$.
One can observe the similarity of Z defined by
 (5) and  (1), if identify
 $w$ with  $i$,
$\phi(w)$ with $E_i$, $\alpha$
with $-\beta$, $\mu=-\frac{Q}{\alpha}$ and
$z=\int d^2w\sqrt{\hat g}e^{\alpha\phi(w)}$
resembles   $\sum_ie^{-\beta E_i}$. In  (1) there is a normal distribution like  
(5), the main difference -in (1) the normal distribution is non-diagonal.
In case of REM we have $2^N$ physical degrees of freedom, like $(L/a)^d$  
degrees in a field theoretical model with ultraviolet $a$ and infrared $L$ cutoffs. 
The ensemble average (integration with a normal distribution of energies) of partition 
function's $\mu$-th  degree corresponds to our expression (1).

In the section 2 we are going to introduce directed polymer(DP)  model on the 
hierarchic trees with branching number q. The endpoints of the hierarchic tree correspond 
to the points $w_i$ of the 2-d space in (1). The case $q\to 1$ resembles the model (1) 
(for a field theoretical aspects see [17]), 
and the case $q\to \infty$ is equivalent to the REM.
There is a strict result [5] that at the case $\mu-> 0$ the thermodynamic 
limit of the introduced models are independent of q.  In the section 3
we give a qualitative derivation of REM solution at complex temperature
 and replica numbers. In section 4 we give the classification of the phase structure of 
the model (1).
In the appendix B. we prove, that in the opposite case $q\to \infty$ DP is equivalent to REM
 the thermodynamic limit. In Appendix A there is a rigorous
 solution of REM at complex temperatures and replica numbers. \\

\section{Hierarchic trees with constant branching number q}

Let us consider the model on the hierarchic tree [2-3],[5].
Originally one has a point (origin of the tree). At the first level of hierarchy there are 
$q$ branches. At the $i$-th  level of hierarchy there are $q$ new branches from the every branch of the $i-1$-th 
 level. At the last K-th level we have $q^K$ end points. Let 
 us consider field $\phi(x)$ at the endpoints. 
  Every point $x$ is connected with the origin of the tree  with a single path. For the any 
  pair of points $x$ and $x'$ at the level $K$ it is possible introduce a hierarchic distance
 \begin{equation}
\label{e6}
v(x,x')=\frac{(K-i)V}{K},
\end{equation}
 where their paths to the origin meet at the $i$-th  level  of hierarchy, V is a parameter 
(the maximal hierarchic distance between points on   the tree). \\
We define random variables $f_{il}$ on the branches at the $i$-th level of the
 tree with distribution
\begin{eqnarray}
\label{e7}
\sqrt{\frac{K}{2V\pi}}\exp\{-\frac{K}{2V}f_{il}^2\}
\end{eqnarray}
We define fields $\phi(x)$ as a sum of $f_{il}$ along the path $il(x)$ 
connecting 
the point $x$ with the origin:
\begin{eqnarray}
\label{e8}
\phi(x)=\sum_{il(x)}f_{il}
\end{eqnarray}
 One can check, that
\begin{equation}
\label{e9}
<\phi(x)\phi(x')>=V-v(x,x')
\end{equation}
If one defines the distance between two points $x,x'$ as
\begin{equation}
\label{e10}
r(x,x')^2=\exp(v(x,x')),
\end{equation}
then Eq.(10) coincides with the ordinary expression of the 2d free field with the 
action (2)
\begin{equation}
\label{e11}
<\phi(x)\phi(x')>=\ln \frac{L^2}{r^2}
\end{equation}
with ultraviolet cutoff $L=\exp(V/2)$ and infrared one $1$.\\
Actually we are interested only in the distance $r(x,x')$ for $x\ne x'$ (for the 
$x'=x$ we can take $r(x,x)=0$). 

We can construct a model connected with the one defined by Eq. (1).
Let us consider a partition
\begin{eqnarray}
\label{e12}
<[\exp\{\sum_x\phi(x)\}]^{\mu}>.
\end{eqnarray}
At the limit  $\mu\to 0$ this model  has been considered rigorously in [5].
At the thermodynamic limit the model is
equivalent to REM (with the same total number  of configurations $q^K$ and variance 
$<E^2>=<\phi(x)^2>$) for all values of $q$.

We suggest the first hypothesis of this work (it can be checked numerically), 
{\bf the thermodynamic limit of Eq.(12) is independent of parameter q like the case 
$\mu\to 0$}.

 We can find the phase structure of the model at the limit $q\to \infty$, it is
  again  equivalent to REM.
While the system (6)-(7),(12) is similar to the system (1) at every value of q due to 
property (9)-(11), there is a serious difference between different choice of q. 
In the case of finite $q$ one should consider a combinatorial problem. Only the limiting 
case $q\to 1$ is similar to 2d Euclidean space, as one can use a small parameter $q-1$ 
and construct measure. Thus we formulate the second hypothesis of the work:
{\bf at the limit $q\to 1$ model on hierarchic tree is equivalent to the system 
(1)}.

Let us define expression (12) for the general case of $q$, then take the limit 
$q\to 1$. 
First we introduce the $\delta $ function in integral representation for the partition 
$z\equiv \exp\{\alpha\sum_x\phi(x)\}$ and Eq. (12) transforms into:
\begin{eqnarray}
\label{e13}
<z^{\mu}>=\int_{\infty}^{\infty}dudv\delta (Re z-u)
\delta (Im z-v)(u+iv)^{\mu}=\nonumber\\
\frac{1}{(2\pi)^2}\int_{-\infty}^{\infty}dk_1dk_2\int_{\infty}^{\infty}dudv(u+iv)^{\mu}\exp(-ik_1u-ik_2v)
G(k_1,k_2)\nonumber\\
G(k_1,k_2)=\nonumber\\
\exp[ik_1Re 
\exp(\alpha\sum_x\phi(x))+ik_2 Im \exp(\alpha\sum_x\phi(x))]
\end{eqnarray}
Now the problem is to calculate the generating function $G(k_1,k_2)$. It can be 
done by 
means of recurrence equations:
\begin{eqnarray}
\label{e14}
I_1(x)=\sqrt{\frac{K}{2V\pi}}\int _{-\infty}^{\infty}\exp\{
-\frac{K}{2V} y^2+U(x+y)\}{\it d }y\nonumber\\
I_{i+1}(x)=\sqrt{\frac{K}{2V\pi}}\int _{-\infty}^{\infty}
\exp\{-\frac{K}{2V} y^2\}
[I_{i}(x+y)]^{q}{\it d }y\nonumber\\
U(y)=ik_1 Re\exp(\alpha y)+ik_2Im\exp(\alpha y)\nonumber\\
G(k_1,k_2)=[I_{K}(0)]^{q}
\end{eqnarray}
Let us consider the limit
\begin{eqnarray}
\label{e15}
q\to 1\qquad K\to \infty\nonumber\\
q^K=\exp(V)
\end{eqnarray}
Then one can express $G(k_1,k_2)$ by means of a function $W(t,x)$:
\begin{eqnarray}
\label{e16}
\frac{d W}{d t}=W\ln W +\frac{1}{2}\frac {d^2W}{dx^2}\nonumber\\
0<t<V, -\infty< x< \infty\nonumber\\
W(0,x)=e^{ik_1 Re\exp(\alpha x)+ik_2Im\exp(\alpha x)}\nonumber\\
G(k_1,k_2)=W(V,0)
\end{eqnarray}
The case of $q\to 1$ trees (13)-(16) with real potential has been considered recently
 in [17]. Using Eq. (16) it is possible to found the phase structure of corresponding 
model in 2d Euclidean space (see. [17]). The suggested method gives an exact phase
 structure (mean field approach gives a correct list of phases but approximate borders
between phases), as well as correct two point correlators and three point corellators
 for isosceles triangles [17]. In this work we are restricted only by phase structure.

In principle one can find numerical solution of the last system (16) and compare it
with the REM's analytical results for the free energy to check the first hypothesis of the work ($q$ 
independence of thermodynamic limit for the model on a $q$-tree). In the next section 
we find the phase structure of REM,  then in the appendix we prove, that system (14) at
 large $q$ is equivalent to REM.\\

\section{Qualitative derivation of 4 REM phases.}

Our goal is to calculate
\begin{eqnarray}
\label{e17}
Z=<z^{\mu_1+i\mu_2}>,\quad
z=\sum_ie^{-(\beta_1+i\beta_2)E_i}
\end{eqnarray}
where energies are distributed via (3).
Let us consider these expressions
for positive integer values of $\mu$,
where the average is over the distribution (3) for each $E_i$.
There are  two competing terms in expression of $z^{\mu}$ (after series expansion).\\
The paramagnetic (PM) phase is originated from the
cross terms in the $z^{\mu}$ series expansion expansion:
\begin{eqnarray}
\label{e18}
Z=
M^{\mu}<e^{-\beta E_{i_1}-\beta E_{i_2}-..
\beta E_{i_{\mu}}}>\nonumber\\
\ln Z=\mu\ln M+N\frac{\beta^2\mu}{2}=
N\frac{(\beta_c^2+\beta^2)\mu}{2}\nonumber\\
N\beta_c^2=2\ln M
\end{eqnarray}
The second one is the correlated paramagnetic (CPM) [4] (in Parisi's picture there 
is a correlation between different replicas), it is originated from the diagonal 
terms in the $z^{\mu}$ series expansion like  to $e^{-\beta{\mu} E_i}$:
\begin{eqnarray}
\label{e19}
Z=<(\sum_{i=1}^Me^{-{\mu}\beta E_i})>\nonumber\\
\ln Z=\ln M+\frac{N\beta^2\mu^2}{2}=\frac{N(\beta_c^2+
\beta^2\mu^2)}{2}.
\end{eqnarray}
Let us consider continuation of (18) to the region $\mu<1$. At critical
temperature $\beta_c$ it's  entropy $\ln Z-\beta\frac{d \ln Z}{d \beta}$ disappears. 
We assume that in this region $\ln Z$ is proportional to $\beta$
(it is natural for a system with zero entropy) and $\mu$.
The continuity of $\ln Z$ gives for spin-glass (SG) phase
\begin{eqnarray}
\label{e20}
\ln Z=N\mu\beta_c\beta
\end{eqnarray}
If one goes to complex temperatures [11-12], then (18) transforms to (it is easy check 
directly for integer $\mu$)
\begin{eqnarray}
\label{e21}
\ln Z=N\frac{(\beta_c^2+\beta_1^2-\beta_2^2)\mu}{2}
\end{eqnarray}
For the SG phase one has to replace $\beta$ by $\beta_1$ in (20): 
\begin{eqnarray}
\label{e22}
\ln Z=N\mu\beta_c\beta_1
\end{eqnarray}
For complex temperatures there is a fourth,
Lee-Yang-Fisher (LYF) phase. The derivation is not direct. The point is,
that for noninteger values of $\mu$
\begin{eqnarray}
\label{e23}
Z\sim <|z|^{\mu}>
\end{eqnarray}
After this trick it is easy to derive the LYF expression. The principal terms are 
$e^{-2\beta_1 E_i}$:
\begin{eqnarray}
\label{e24}
\ln Z=\frac{N(\beta_c^2+4\beta_1^2)\mu}{4}
\end{eqnarray}
Let us now continue  our four expressions to complex values of $\mu$.
For PM phase an analytical continuation of Eq. (18) gives
\begin{eqnarray}
\label{e25}
\ln Z=N\frac{(\beta_c^2+\beta_1^2-\beta_2^2)\mu_1-2\beta_1\beta_2\mu_2}{2}
\end{eqnarray}
For SG phase we have
\begin{eqnarray}
\label{e26}
\ln Z=N\mu_1\beta_c\beta_1
\end{eqnarray}
For LYF phase:
\begin{eqnarray}
\label{e27}
\ln Z=\frac{N(\beta_c^2+4\beta_1^2)\mu_1}{4}
\end{eqnarray}
For CPM an analytical continuation of Eq. (19) gives
\begin{eqnarray}
\label{e28}
\ln Z=\frac{N[\beta_c^2+(\beta_1^2-\beta_2^2)(\mu_1^2-\mu_2^2)-4\beta_1\beta_2\mu_1\mu_2)]}{2}
\end{eqnarray}
To find the borders between four phases one should first find
the correct phase at $\mu\to 0$ limit, then compare its finite $\mu$ expression
for $|\ln Z|$ with the corresponding one given by CPM phase. It is known, that LYF phase
exists at [14,15]:
\begin{eqnarray}
\label{e29}
\beta<\frac{\beta_c}{2}
\end{eqnarray}
 and PM one at
$\beta<\beta_c$.
For a complex temperatures one has a condition for SG phase
\begin{eqnarray}
\label{e30}
\beta_1>\beta_c+\beta_2.
\end{eqnarray}
The last point. Strict derivation gives, that LYF for noninteger $\mu_1$ exists only
 at
 \begin{eqnarray}
\label{e31}
\mu_1>-2
\end{eqnarray}
The paramagnetic phase is the most symmetric one, there are local symmetries in the
 model.\\
For the case of SG phase there is some order and no local symmetry. For the case
of LYF phase there is some correlation between couples of replicas. Those two phases
 (SG
 and LYF) resemble non-unitary models in field theory. For the case of CPM
phase there is a correlation between all the replicas, but local symmetries are conserved.
This phase has not any pathology.
Thus together with PM phase it can be connected with 
unitary models.

\section{String phases}
What can we say about quantum field theory and strings on the basis of our results?
One should understand that, in some sense, there is a hierarchy of requriements 
physical theory to be mathematically rigorous.

On the top level in quantum field theory one demands the unitarity of the theory as the main constraint.
In the lower level of statistical field theory in Euclidean space the genesis of two probabilities:
in ensemble and Boltzmann  one is not too complicated. Therefore one can consider non-unitary model also,
connected with some physical situations [18]. We have another physical constraint: 
finite number of primary fields.
In the lowest level of hierarchy is an ordinary statistical mechanics. Here one is happy, when can 
construct a thermodynamic limit ($N\to \infty$) for the free energy, entropy. 

We see, that our results are quite restrictive.
First, we should forbid a situation like Lee-Yang-Fisher singularity with too negative
replica numbers $\mu_1<-2$. Here it is impossible construct a thermodynamic limit.
Other situations with Lee-Yang-Fisher phases as well as with spin-glass are a bit interesting,
but here could not be any unitary theory.
The most interesting are paramagnetic and correlated paramagnetic phases. In this area
there is a some chance for unitarity.

Let us return to the partition of bosonic d-dimensional string (1). For the ultraviolet
cutoff $L$ and infrared one $a$ the number of degrees is 
\begin{eqnarray}
\label{e32}
M=\frac{L^2}{a^2}.
\end{eqnarray}
configurations.
Let us define distribution of $\phi(w)$ over all points $w$,
using the free field action from (1):
\begin{eqnarray}
\label{e33}
\rho(\phi_0)\equiv <\delta(\phi_0-\phi(w)>_{\phi(w)} \sim
\exp(-\frac{\phi_0^2}{2G(0)}),
\end{eqnarray}
where $G$ is correlator of $\phi(w)$ fields,
the average is over the  distribution
\begin{eqnarray}
\label{e34}
\rho(\phi(w))\sim e^{\frac
{1}{8\pi}\int d^2w\sqrt{\hat g}{\phi \Delta \phi +QR\phi}},
\end{eqnarray}
and
\begin{eqnarray}
\label{e35}
G(0)=2\ln \frac{L}{a}.
\end{eqnarray}
We replace our system (1) with a REM model having the same number $M$ independent variables $E_i\sim 
\phi(w)$
with the same distribution (32):
\begin{eqnarray}
\label{e36}
N=G(0),\alpha_c=\sqrt{\frac{2\ln M}{G(0)}}=\sqrt{2}\nonumber\\
\ln Z_{PM}=2\frac{\alpha^2+\alpha_c^2}{2}\ln\frac{L}{a}
\end{eqnarray}
Here $Z_{PM}$ is the partition connected with the Eq. (1) for the PM phase at real $\alpha$.
 We rescale the temperature:
\begin{eqnarray}
\label{e37}
\frac{\alpha}{\sqrt{2}}=\beta, \beta_c=1,\mu=-\frac{Q}{\alpha}
\end{eqnarray}
DDK formulas give for $d\equiv c$ dimensions:
\begin{eqnarray}
\label{e38}
Q=\sqrt{\frac{25-c}{3}},\alpha=-\frac{1}{\sqrt{12}}(\sqrt{25-c}-\sqrt{1-c})
\end{eqnarray}
For the sphere topology:
For the $1<d<25$:
\begin{eqnarray}
\label{e39}
\beta_1=\frac{\sqrt{25-c}}{\sqrt{24}},\beta_2=-\frac{\sqrt{c-1}}{
\sqrt{24}}\nonumber\\
\mu_1=\frac{1}{12}[25-c], \mu_2=\sqrt{(25-c)(c-1)}\frac{1}{12}
\end{eqnarray}
For $25<d<26$ we have:
\begin{eqnarray}
\label{e40}
\beta_2=\frac{\sqrt{c-25}}{\sqrt{24}}-\frac{\sqrt{c-1}}{
\sqrt{24}}\nonumber\\
\mu_1=\frac{1}{12}[25-c]+\sqrt{(25-c)(c-1)}\frac{1}{12}
\end{eqnarray}
For other string topologies one should rescale the $\mu$ expressions in Eqs. (39)-(40):
\begin{eqnarray}
\label{e41}
\mu\to (1-g)\mu
\end{eqnarray}

Let us consider first the sphere topology.
We denote $y=\frac{25-d}{24}$. For the $1<d<25$ we have:\\
$\beta_1=\sqrt{y},\beta_2=-\sqrt{1-y},\mu_1=2y,\mu_2=2\sqrt{y(1-y)}$.
We derfive the following 4 expressions for the $\ln Z$:
\begin{eqnarray}
\label{e42}
\frac{\mu_1[1+(\beta_1^2-\beta_2^2)]-2\mu_2\beta_1\beta_2}{2}=\frac{4y^2+4y(1-y)}{2}=2y,PM\nonumber\\
\frac{1+(\mu_1^2-\mu_2^2)(\beta_1^2-\beta_2^2)-4\mu_1\mu_2\beta_1\beta_2}{2}=
\frac{1+4y(2y-1)^2+16y^2(1-y)}{2}=\frac{1+4y}{2},CPM\nonumber\\
\mu_1\beta_1=2y^{3/2},SG\nonumber\\
\frac{\mu_1(1+4\beta_1^2)}{4}=\frac{y(1+4y)}{2},LYF
\end{eqnarray}
We see, that CPM phase is prefferable in the region $1\le d \le19$ ($0\le y\le 1$).\\
For the $19\le d\le26$ we should compare $\ln Z$ expressions for the CPM and LYF phases,
as $\beta_1=0,\mu_1>0$. Now we denotee $y=-\frac{25-d}{24},\mu_1=-2y+2\sqrt{y(y+1)},
\beta_2=y-\sqrt{1+y}$. For the CPM phase we have 
$\ln Z=\frac{1-\mu_1^2\beta_2^2}{2}=(1-4(y-\sqrt{y(y+1)})^2(y-\sqrt{1+y})^2/4$ and for the LYF
phase $\ln Z=\frac{\mu_1}{4}=\frac{-2y+2\sqrt{y+1}}{4}$.
We see that for spherical topology for the whole region $1\le d\le 26$ string is in CPM phase.

Let us consider the torus topolgy case. We have for PM phase $\ln Z=\frac{1+(\beta_1^2-\beta_2^2)}{2}\equiv y$,
for SG phase $\ln Z=\beta_1=\sqrt{y}$ and for LYF phase $\ln Z=\frac{1+\beta_1^2}{4}=\frac{1+4y}{4}$, where
$y=\frac{25-d}{24}$.
At $1\le d\le 19$ system with torus 
topology is in the SG phase. LYF phase exists  at  $19\le d\le 26$. 

Let us consider now now higher topolgies $g\le 2$.
Again system is in SG phase for $1\le d\le 19$. At $19<d<25$ still exists a thermodynamic limit
and system is in LYF phase, if $\frac{25-d}{4}(g-1)<2$, therefore $g=5, d=19$ is a multicritic point.
For the $d=26$ there is a thermodynamic limit with LYF phase at
$(g-1)/3<2$. 

Let us consider now a case of superstring. Now one has:
$$Q=\sqrt{\frac{9-d}{2}},-\alpha=\frac{\sqrt{9-d}-i\sqrt{c-1}}{2\sqrt{2}}$$
Let us denote $u=\frac{9-d}{8}$
$$\beta_1=\frac{\sqrt{9-c}}{2\sqrt{2}}=u^{1/2},\beta_2=-\frac{\sqrt{d-1}}{2\sqrt{2}}=-(1-u)^{1/2}$$
$$\mu_1=\frac{9-d}{4}=2u,\mu_2=\frac{\sqrt{(9-d)(1-d)}}{4}=2\sqrt{u(1-u)}$$
We see a mapping $y\to u$. Now the transition to the LYF phase is at $d=7$.
According the [19] interesting dimension is $d=5$, connected with QCD interpretation as strings [20].

What one can say about string's physics on the ground of the REM picture?
The most interesting case is the sphere case. When one climbs over the $d=1$ barrier, nothing happens in REM picture, system is still in CPM phase, 
as for the $d<1$. The free energy has not any singularity (might be there are singularities in some correlators).
To reveal
interesting (unitary) theories explicitly one should solve the
directed polymer at finite replica number including finite size corrections and correlators.
 But at least for the sphere case the REM analysis
seems to be quite reliable.

\section{Conclusions}

In sections 1,2 and Appendix B we gave an arguments for the connection of string's
partition with a finite replica number REM.
In section 3 and in Appendix A we solved Random Energy Modela at complex temperatures and replica numbers.
In section 4 we take string model with  an analytical continuation of
the David-Kawai-Distler formulas at $d>1$ and mapped it to REM. 
The validity of DDK formulas at
 $d>1$ is still under question, but we hope that the analytical
continuation could reveal singularities of the system. It is a typical situation
in statistical physics, when there is a singularity in free energy expression, when it is
analytically continued from one of phases to the border between phases. 

We obtained a bit strange result about difference of phases
for the different topologies of string surfaces at $d>1$. For the spherical case
there is no any barrier at $d=1$, at least for the free energy.
For the other topolgies at $d>1$ system is in SG or LYF phase, sometimes 
the model is so pathalogical thet there is no any thermodynamic limit.

There have been early attempts to connect strings with spin glasses.
I have several discussions with V. Knizhnik in Alma-Ata conference
in 1985, later in Yerevan before his death. He was highly intrigued
with a ultrametry property of spin-glasses and trying to connect them with string.

M. Virassoro also informed me about his and G. Parisi's attempts to connect strings with 
spin glasses.

 We could succeed due  to work done in [4], [6-8] and a simple observation 
     that string's partition is similar to finite replica
    REM just after zero mode integration.
     String theory is too mathematized. In this work we tried to  catch
     some narrow but crucial aspect of the theory using more physics
     and less complicated mathematical tools.

I am grateful to B.Derrida, Y. Sinai for the discussion of $q=1$ trees.

\renewcommand{\theequation}{A.\arabic{equation}}

\setcounter{equation}{0}
\appendix{}
\section{REM's solution for complex temperatures and real replica number.}

To calculate expression (5) we introduce an identity 
$$\int dU_1\delta(U_1-Re \sum_ie^{(\beta_1+i\beta_2)E_i}
\int dU_2\delta(U_1-Im \sum_ie^{(\beta_1+i\beta_2)E_i}=1$$
and an integral representation for 
$\delta$ function $\delta(z-u)\equiv \frac{1}{2\pi}\int dke^{ik(z-u)}$:
\begin{eqnarray}
\label{a1}
f(k_1,k_2)\equiv g(k_1,k_2)^M=\nonumber\\
\frac{1}{\sqrt{N\pi}}
\int_{-\infty}^{\infty}dx
\exp[\frac{-x^2}{2N}]\exp(ik_1e^{\beta_1x}\cos(\beta_2x)+
ik_2e^{\beta_1x}sin(\beta_2x))]^M
\end{eqnarray}
Having an expression for the function $f(k1,k2)$ we can define the partition $Z$:
\begin{eqnarray}
\label{a2}
Z=\nonumber\\
\frac{1}{4\pi^2}\int_{-\infty}^{\infty}dk_1dk_2dU_1dU_2
e^{-ik_1U_1-ik_2U_2}(U_1+iU_2)^{\mu}f(k_1,k_2)
\end{eqnarray}
This is an exact expression. In thermodynamic limit we will consider four different 
asymptotics for the function $f(k1,k2)$.

In the paramagnetic phase  we expand an exponent via degrees of $k_1,k_2$:
\begin{eqnarray}
\label{a3}
g(k_1,k_2)\approx 1+ik_1Re e^{N\frac{(\beta_1^2-\beta_2^2)+i2\beta_1\beta_2}{2}}+ik_2Im e^{N\frac{
(\beta_1^2-\beta_2^2)+i2\beta_1\beta_2}{2}}
\end{eqnarray}
Integration via $dk_1,dk_2$ gives 
$$\delta(U_1-Re e^{N\frac{
(\beta_1^2-\beta_2^2)+i2\beta_1\beta_2}{2}})\delta(U_2-Im e^{N\frac{
(\beta_1^2-\beta_2^2)+i2\beta_1\beta_2}{2}}).$$ 
Eventually we  derive for the PM phase:
\begin{eqnarray}
\label{a4}
f(k_1,k_2)\approx\exp[ik_1M Re e^{N\frac{
(\beta_1^2-\beta_2^2)+i2\beta_1\beta_2}{2}}+ik_2M Im e^{N\frac{
(\beta_1^2-\beta_2^2)+i2\beta_1\beta_2}{2}}]\nonumber\\
\ln <z^{\mu}>=N\frac{\mu_1(\beta_c^2+\beta_1^2-\beta_2^2)-2\mu_2\beta_1\beta_2}{2}
\end{eqnarray}
We miss the imaginary part in the expression of $\ln <Z>$.

For the Lee-Yang-Fisher (LYF) phase we take the second terms in the expansion of the
exponent :
\begin{eqnarray}
\label{a5}
g(k_1,k_2)\approx 1-\frac{1}{\sqrt{2N\pi}}\int_{\infty}^{\infty}e^{\frac{-x^2}{2N}+
2\beta_1x}\frac{(k_1\cos(\beta_2x)+k_2sin(\beta_2x))^2}{2}\nonumber\\
=1-\frac{k_1^2+k_2^2}{4}e^{2N\beta_1^2}
\end{eqnarray}
We obtain:
\begin{eqnarray}
\label{a6}
f(k_1,k_2)\approx
\exp[-M\frac{k_1^2+k_2^2}{4}e^{2N\beta_1^2}]\nonumber\\
Z=\nonumber\\
\frac{1}{4\pi^2}\int_{\infty}^{\infty}dk_1dk_2dU_1dU_2e^{-ik_1U_1-ik_2U_2}(U_1+iU_2)^{\mu}\frac{k_1^2+k_2^2}{4}\exp[-Me^{2N\beta_1^2}]\nonumber\\
=\frac{1}{\pi Me^{2N\beta_1^2}}\int dU_1dU_2\exp[-\frac{(U_1^2+U_2^2)}{Me^{2N\beta_1^2}}](U_1+iU_2)^{\mu_1+i\mu_2}\nonumber\\
=e^{\mu_1N\frac{\beta_c^2+4\beta_1^2}{4}}\frac{1}{\pi}\int_{0}^{\infty} dr\int_0^{2\pi}d{\varphi}\exp[-r^2]r^{\mu_1+1+i\mu_2}e^{(\mu_1+i\mu_2)i\varphi}\nonumber\\
=\frac{1}{\pi}`e^{\mu_1N\frac{\beta_c^2+4\beta_1^2}{4}}\Gamma(\frac{\mu_1+1+i\mu_2}{2})\frac{\exp(2\pi(\mu_1+i\mu_2))-1}{\mu_1+i\mu_2}
\end{eqnarray}
The LYF phase (A5) is rigorously defined for $\mu_1>-2$, otherwise there is a singularity at
integration $dU_1dU_2$.
 
For the CPM phase we consider the case of positive  $\mu_1$. We take   $n>\mu_1>n-1,\nu\equiv\mu_1-n, 0>\nu>-1$
and write an equivalent expression for the (A1):
\begin{eqnarray}
\label{a7}
Z=
\frac{1}{4\pi^2}\int_{-\infty}^{\infty}dk_1dk_2 f(k_1,k_2)
\int_{-\infty}^{\infty}dU_1dU_2
(\frac{id}{dk_1}-\frac{d}{dk_2})^ne^{-ik_1U_1-ik_2U_2}\nonumber\\
(U_1+iU_2)^{\nu}
\end{eqnarray}
Now we assume, that in the principal region of ${\it d}k_1,{\it d}k_2$ integration
\begin{equation}
\label{a8}
f(k_1,k_2)-1\ll 1.
\end{equation}
Therefore we can expand the exponent in the $f(k_1,k_2)$ expression: 
$$f(k_1,k_2)=[\frac{1}{\sqrt{\pi N}}\int_{\infty}^{\infty}dx
\exp[-\frac{x^2}{N}+ik_1\cos(\beta_2x)+ik_2sin(\beta_2x)]]^M$$
$$\approx 1+M\{\int_{-\infty}^{\infty}\frac{dk_1dk_2}{\sqrt{\pi N}}dx
\exp[-\frac{x^2}{N}+ie^{\beta_1x}(k_1\cos(\beta_2x)+
k_2sin(\beta_2x))]-1\}$$
Then after integration by parts:
\begin{eqnarray}
\label{a9}
Z=\nonumber\\
\frac{M}{4\pi^2}\int_{-\infty}^{\infty}dk_1dk_2dU_1dU_2
e^{-ik_1U_1-ikU_2}(U_1+iU_2)^{\nu}
(-\frac{id}{dk_1}+\frac{d}{dk_2})^nf(k_1,k_2)\nonumber\\
=\frac{M}{4\pi^2}\int_{-\infty}^{\infty}dk_1dk_2\int_{-\infty}
^{\infty}dU_1dU_2e^{-ik_1U_1-ikU_2}(U_1+iU_2)^{\nu}
\frac{1}{\sqrt{\pi N}}dx \nonumber\\
\exp[-\frac{x^2}{N}+iRe(k_1-ik_2)e^{(\beta_1+i\beta_2)x}]
e^{(\beta_1+i\beta_2)xn}
\end{eqnarray}
We miss the term $1$ in the expression of $f(k_1,k_2)$, because its
 contribution is equal to $0$ after integration by parts.
Let us denote     $E=\exp[(\beta_1+i\beta_2)x],K=k\exp(i\varphi)=k_1+ik_2,U=U_1+iU_2$.
First we take the integration via $dk_1,dk_2$. The result is
$\delta(E-U)$. Then we calculate Gaussian integral via $dx$ and derive an expression
for the correlated paramagnetic phase (CPM):
\begin{eqnarray}
\label{a10}
Z\equiv<z^{\mu}>=M\frac{1}{\sqrt{\pi
N}}\int_{\infty}^{\infty}{\it d}xe^{-\frac{x^2}{N}+
\mu x(\beta_1+i\beta_2)x}=\nonumber\\
\exp[N\frac{(\mu_1^2-\mu_2^2)
(\beta_1^2-\beta_2^2)-4\beta_1\beta_2\mu_1\mu_2+\beta_c^2}{2}]
\end{eqnarray}
Let us calculate SG phase. It is convinent to use another
representation of function $f(k_1,k_2)$ [16].
Using the  Stratanovich transformation for the energy density term 
$$\exp(-\frac{x^2}{2})=\frac{\beta_1\sqrt{N}}{\sqrt{2\pi}}
\int_{-i\infty}^{i\infty}{\it d}ye^{\frac{N\beta_1^2y^2}{2}+\beta_1\sqrt{N}xy}$$
we derive
\begin{eqnarray}
\label{a11}
g(k_1,k_2)=
\frac{\sqrt{N}}{2\pi}\int_{\infty}^{\infty}dy\int_{-\infty}^{\infty}dx
\exp(\frac{N\beta_1^2y^2}{2}+
\beta_1\sqrt{N}xy+ik_1Ree^{(\beta_1+i\beta_2)\sqrt{N}x}+
ik_2Ime^{(\beta_1+i\beta_2)\sqrt{N}x})
\end{eqnarray}
After transformation $v=k\exp(\sqrt{N}\beta_1x)$ we have 
\begin{eqnarray}
\label{a12}
f(k_1,k_2)=(\frac{1}{2\pi i}\int_{-i\infty}^{i\infty}{\it d}y
e^{-y \ln k+\frac{N\beta_1^2y^2}{2}}G(y,k,\varphi))\nonumber\\
G(y,k,\varphi)=\int_{0}^{\infty}{\it d}v
e^{iv \cos[\beta_2/\beta_1(\ln v-\ln k)-\varphi]}v^{(y-1)}.
\end{eqnarray}
We are interesting in Eq.(12) for the $|\ln k|\sim N$, therefore we can calculate the asymptotic
of the function $f(k1,k2)$ via the saddle point method.
There is a pole of function $G(y)$ at  $y=0$ with residue equal to 1 (it can be 
derived putting a smal low integration limit). Let us shift 
the integration loop to the saddle point. We have:
\begin{equation}
\label{a13}
f(k,\varphi)=M(1+\frac{1}{2\pi i}\int_{-i\infty}^{i\infty}{\it d}y
e^{- y\ln |k|}e^{N\frac{\beta_1^2 y^2}{2}}G(y,k,\varphi))
\end{equation}
For the saddle point we have
\begin{equation}
\label{a14}
y_0=\frac{\ln k}{N\beta_1^2}
\end{equation}
We move the integration loop via $dy$ to catch the saddle point. 
For the analytical continuation to the region $[-1<Re y<0]$ we transform
expression of $G(y)$ from (A12) using the integration by parts:
\begin{eqnarray}
\label{a15}
G(y,k,\varphi)=-i\int_{0}^{\infty}{\it d}vv^{y}
\exp[i v \cos[\beta_2/\beta_1(\ln \frac{v}{k}-\varphi]]\{\cos[\beta_2/\beta_1(\ln \frac{v}{k}-\varphi]-\beta_2/\beta_1\sin
[\beta_2/\beta_1(\ln \frac{v}{k}-\varphi]\}
\end{eqnarray}
We have an asymptotics:
\begin{eqnarray}
\label{a16}
g(k,\varphi)=1-\frac{1}{\sqrt{\pi N}\beta_1}[
-G(\frac{2\ln k}{N\beta_1^2},k,\varphi)]e^{-\frac{\ln k^2}
{N\beta_1^2}}\nonumber\\
f(k_1,k_2)=exp[-Me^{-\frac{\ln k^2}
{N\beta_1^2}}A]\nonumber\\
A=-G(y_0,k,\varphi)
\end{eqnarray}
One should take only this asymptotic instead of (A3),(A5) if, while  shifting the integration loop,
we don't intersect the pole at $y=-1$,
\begin{equation}
\label{a17}
\frac{|\ln k|}{N\beta_1^2}<1
\end{equation}
Otherwise, at $\frac{|\ln k|}{N\beta_1^2}>1$ we should consider all three different asymptotics
 (A3),(A5),(A16) and choose the largest one.
From the Eq. (a16) we derive immediatly the bulk value of partition:
\begin{eqnarray}
\label{a18}
Z\sim \exp[\mu_1N\beta_1\beta_c]
\end{eqnarray}
We derived accurate expressions for the PM phase (A4),LYF phase (A6), CPM phase (A10)
and bulk expression for the SG phase (A17).
We see, that LYF phase can be constructed only at $\mu_1>-2$. Thus a situation, when bulk expression
for the $<Z>$ is given by LYF phase and $\mu_1<-2$, model is to pathalogic and there is no thermodynamic limit.

To find the borders between phases one should solve the model at the limit $\mu\to 0$,
then compare the corresponding expression of the largest free energy with the one in CPM for the finite 
$\mu_1$ (it exists only at
 $\mu_1>0$). One should choose a phase, having larger 
value of  $|\ln Z|$.
Let us remember also that SG phase exists
at $\beta_1+\beta_2>\beta_c$,
the LYF phase at $\beta_2>\frac{\beta_c}{2}, \beta_1<\frac{\beta_c}
{2}$.

\section{\large  Equivalence of REM and directed polymer in 
thermodynamic limit}
\renewcommand{\theequation}{B.\arabic{equation}}

\setcounter{equation}{0}

We consider a hierarchic tree with $K$ levels and large $Q$.
We have, that along any path connecting endpoint with the origin:
\begin{eqnarray}
\label{b1}
\sum_{\alpha}<\epsilon_{\alpha}^2>=N
\end{eqnarray}
Let us first consider the PM phase. We should define the generating function
\begin{eqnarray}
\label{b2}
<f(k_1,k_2)>\equiv\nonumber\\ \sum_{}<\exp[iRe (k_1-ik_2)e^{\beta_1+i\beta_2)(\epsilon_{i_1}+
\epsilon_{i_2}+..\epsilon_{i_K})}]>
\end{eqnarray}
Here the sum is over all the paths, connecting endpoints with the origin.
Let us consider first the integration via the last level of hierarchy. We expand an
exponent and after integration via $d\epsilon_{i_K}$:
 \begin{eqnarray}
\label{b3}
<G(k_1,k_2)>=<\exp[iRe(k_1-ik_2)e^{(\beta_1+i\beta_2)(\epsilon_1+\epsilon_2
+..\epsilon_{K-1})}]
Qe^{N\frac{\beta_1^2-\beta_2^2+i\beta_1\beta_2}{2K}}>
\end{eqnarray}
Repeating this procedure K times, we obtain
 \begin{eqnarray}
\label{b4}
<f(k_1,k_2)>=\exp[iRe(k_1-ik_2)Q^Ke^{N\frac{\beta_1^2-\beta_2^2+i2\beta_1\beta_2}{2}}]
\end{eqnarray}
In principle the expression in the exponent could be large.
We recover the REM result for the PM phase with accuracy o(1).\\
For the LYF the integration via the last level of hierarchy gives
 \begin{eqnarray}
\label{b5}
<f(k_1,k_2)>\equiv
<1-\frac{k_1^2+k_2^2}{4}e^{2\beta_1(\epsilon_1+\epsilon_2+..\epsilon_{K-1})}
Q\frac{k_1^2+k_2^2}{4}e^{\frac{N\beta_1^2}{K}}]>
\end{eqnarray}
Repeating the integration K times (we expand the exponent all the time via 
$(k_1^2+k_2^2)$)
gives the REM result for LYF phase
 \begin{eqnarray}
\label{b6}
<f(k_1,k_2)>=\exp[-\frac{k^2_1+k_2^2}{4}Q^Ke^{N{\beta_1^2}}]
\end{eqnarray}
Again we have equivalence with accuracy O(1).\\
The case of SG phase is a bit complicated. Now the integration via the last level of 
hierarchy gives
 \begin{eqnarray}
\label{b7}
<f(k_1,k_2)>=\exp[-\frac{Q}{\sqrt{N}c}e^{-\frac{[\ln k +\beta_1(\epsilon_1+
\epsilon_2+..\epsilon_{K-1})]^2}{N\beta_1^2}}]\nonumber\\
k=\sqrt{k_1^2+k_2^2}
\end{eqnarray}
where $c\quad O(1)$. This expression resembles the case of real temperatures, where the
 generating function is calculated [16].  Using those results for the real temperature
directed polymer, we derive again the REM expression for the SG phase.

Let us consider now the case of CPM phase. Here we use again formulas (a7),
(a8).
Now all the integrations decouple and we recover the result of a simple REM:
 \begin{eqnarray}
\label{b8}
<z^{\mu}>=M<e^{-(\mu_1+i\mu_2)(\beta_1+i\beta_2)x}>=\nonumber\\
\exp[N\frac{(\mu_1^2-\mu_2^2)
(\beta_1^2-\beta_2^2)-4\beta_1\beta_2\mu_1\mu_2+\beta_c^2}{2}]
\end{eqnarray}


\begin{thebibliography}{99}
\bibitem{b1} B. Derrida, Phys. Rev. Lett. {\bf 45} (1980) 79
\bibitem{b2} B.Derrida,H. Spohn, J. Stat. Phys.  {\bf 51} (1988) 817
\bibitem{b3} J.Cook,B.Derrida,J.Stat.Phys. {\bf 63}(1991)505
\bibitem{b4} E.Gardner,B. Derrida, J.Phys.{\bf A22} (1989)1975
\bibitem{b5} B.Derrida, M.R. Evans,E.R. Speer,Com.Mat. Phys.{\bf15}6(1993)221
\bibitem{b6} C.C.Chamon,C.Mudry,X.G.Wen , Phys. Rev.Lett. {\bf 77}(1996)4194
\bibitem{b7} I.I.Kogan et all Phys. Rev.Lett. {\bf 77}(1996)707
\bibitem{b8} H.E.Castillo et all,Phys.Rev. B.{\bf 56}(1997)10668
\bibitem{b9} A.M. Polyakov Phys. Letters {103B}(1981)207
\bibitem{b10} J. Distler,H.Kawai,Nucl. Phys. {\bf  B321}(1989)504
\bibitem{b11} F. David, Mod. Phys. Lett. {\bf A3}(1988)1651
\bibitem{b12} A. Gupta,S.Trivedi,M.Wise Nuclear Phys. {\it bf B340}(1990)475
\bibitem{b13} E. Abdalla M.C.B.Abdalla,D.Dalmazi,A.Zadra,2d -Gravity in 
Non-critical strings, Springer-Verlag,1994
\bibitem{b14} C. Moukarzel, N. Parga, Physica {\bf A177} (1991) 24
\bibitem{b15} B. Derrida, Physica  {\bf A177} (1991) 31
\bibitem{b16} D. Saakian, Phys.Rev PRevE,{\bf  61},(2000)6132
\bibitem{b17} D. Saakian, Phys.Rev PRevE,{\bf  65}(2002) 67104
\bibitem{b18} J.Cardy, Phys. Rev. Lett. {\bf 54}(1985)1354
\bibitem{b19} J. Polchinski, M. Strasser,J. High Energy Phys. {\bf 05}(2003)
\bibitem{b20}  G. t'Hooft,Nucl.Phys.{\bf 72}(1974)461
\end{thebibliography}
\end{document}